\documentclass{iopart}
\usepackage{iopams}
\newcommand{\eqref}[1]{(\ref{#1})}
\usepackage{amsfonts}

\newcommand{\CH}{\mathrm{CH}}

\newcommand{\vm}{{\boldsymbol{m}}}
\newcommand{\bvm}{{\bar{\vm}}}
\newcommand{\vell}{\boldsymbol{\ell}}
\newcommand{\vn}{\boldsymbol{n}}
\newcommand{\bth}{\boldsymbol{\theta}{}}
\newcommand{\tth}{\tilde{\bth}}
\newcommand{\bom}{\boldsymbol{\omega}}
\newcommand{\bOm}{\boldsymbol{\Omega}}
\newcommand{\ve}{\boldsymbol{e}}
\newcommand{\fo}{\mathfrak{o}}

\newcommand{\vA}{{\boldsymbol{A}}}

\newcommand{\tLambda}{\tilde{\Lambda}}
\newcommand{\ttau}{\tilde{\tau}}
\newcommand{\tmu}{\tilde{\mu}}

\newtheorem{theorem}{Theorem}
\begin{document}

\title{The type N Karlhede bound is sharp} \author{R. Milson, N.
  Pelavas}
\address{Dept. Mathematics and Statistics, Dalhousie University\\
  Halifax NS B3H 3J5, Canada} \eads{\mailto{rmilson@dal.ca},
  \mailto{pelavas@mathstat.dal.ca}}
\begin{abstract} 
  We present a family of four-dimensional Lorentzian manifolds whose
  invariant classification requires the seventh covariant derivative
  of the curvature tensor.  The spacetimes in questions are null
  radiation, type N solutions on an anti-de Sitter background.  The
  large order of the bound is due to the fact that these spacetimes
  are properly $\CH_2$, i.e., curvature homogeneous of order $2$ but
  non-homogeneous. This means that tetrad components of $R, \nabla R,
  \nabla^{(2)}R$ are constant, and that essential coordinates first
  appear as components of $\nabla^{(3)}R$.  Covariant derivatives of
  orders 4,5,6 yield one additional invariant each, and
  $\nabla^{(7)}R$ is needed for invariant classification.  Thus, our
  class proves that the bound of 7 on the order of the covariant
  derivative, first established by Karlhede, is sharp.  Our finding
  corrects an outstanding assertion that invariant classification of
  four-dimensional Lorentzian manifolds requires at most
  $\nabla^{(6)}R$.
\end{abstract}

The equivalence problem for pseudo-Riemannian geometry is of central
importance to general relativity.  The invariant classification (IC)
of a metric requires knowledge of the curvature tensor and its
covariant derivatives $R, \nabla R, \ldots, \nabla^{(q)} R$, up to
sufficiently high order.  The IC order, the smallest value of $q$
required for invariant classification, depends on the spacetime in
question.  In general, spacetimes without isometries (Killing vectors)
and with an algebraically special Weyl tensor require a larger IC
order.  An open problem in general relativity is the maximum IC order
that can occur in the equivalence problem for four-dimensional,
Lorentzian metrics.

Cartan\cite{cartan} was the first to formulate the equivalence problem
in terms of moving frames, structure equations, and differential
invariants, and to provide an initial estimate of $q\leq n(n+1)/2$,
where $n$ is the dimension of the underlying manifold $M$.  Cartan's
ideas were first applied to four-dimensional relativity by
Brans\cite{brans65}.  Subsequently, Karlhede introduced a simplified
algorithm based on the Petrov and Segre classifications of the
curvature tensor \cite{karlhede80}, which provided better estimates for
the IC order.  The algorithm was refined and implemented in a computer
algebra system by MacCallum, \AA man\cite{MacCAman86}, and others; see
\cite{PSdI2000} for a recent review.

Karlhede's counting argument establishes a bound of $q\leq \dim G_0 +
n+1$, where $G_0$ is the isotropy group of $R$.  Using the well-known
Petrov-Penrose classification of the Weyl tensor, the argument yields
$q\leq 5$ for Petrov types I, II, III; $q\leq 7$ for Petrov types N,
D; and $q\leq 8$ for type O\footnote{Here, one has to consider the
  possible symmetries of the Ricci tensor.}.  These bounds have been
improved by careful analysis. It is now known that $q\leq 6$ for type
D \cite{coldinv93}, and that $q\leq 6$ for type O \cite{PS2000}.
However, it is not known whether these type D and type O bounds are
sharp.  Detailed analysis of vacuum type N solutions yields $q\leq 5$
\cite{ramvick96}.  A similar analysis of non-vacuum
type N solutions produced a claim of $q\leq 5$ \cite{ramos98}.  Subsequently,
examples of type N solutions with $q=5$ were discovered
\cite{skea2000}.

However, contrary to the above findings, we exhibit an example of a
type N, non-vacuum solution that has IC order $q=7$, and thereby show
that the type N Karlhede bound is sharp.  In our best estimation, the
apparent discrepancy between our result and previous claims is due by
a subtle error in the analysis of one subcase in \cite{ramos98}.

Let us begin by reviewing the Karlhede IC algorithm.  Let $(M,g)$ be
an $n$-dimensional pseudo-Riemannian manifold.  Let $\eta_{ab}$ be a
constant, non-degenerate quadratic form having the same signature as
the metric $g$. Henceforth, we use $\eta_{ab}$ to raise and lower
frame indices, which we denote by $a,b,c=1,\ldots, n$, and say that a
coframe $\bth^a$ is $\eta$-orthogonal if $g = \eta_{ab} \bth^a
\bth^b$. Let $O(\eta)=\{ (X^a{}_b) : X^a{}_c \eta_{ab} X^b{}_d =
\eta_{cd}\}$ denote the group of $\eta$-orthogonal transformations.
Let $\fo(\eta)=\{ (A^a{}_b): A_{(ab)}=0\}$ denote the
$n(n-1)/2$-dimensional Lie algebra of $\eta$ skew-symmetric
transformations.
\begin{itemize}
\item[1.] Let $q=0, G_{-1} = O(\eta), t_{-1} = 0$.  All
  $\eta$-orthogonal frames are permitted.
\item[2.] Compute $\nabla^{(q)}R$ relative to a permitted
  $\eta$-orthogonal frame. 
\item[3.]  Determine $G_q\subset
  G_{q-1}$, the isotropy group of $\nabla^{(q)}R$.
\item[4.] Restrict the  frame freedom to $G_q$ by putting
  $\nabla^{(q)}R$ into standard form (normalizing some
  components to a constant.) 
\item[5.] The functions in the set $R^q=\{ R_{abcd},
  R_{abcd;e_1},\ldots, R_{abcd;e_1\ldots, e_q} \}$ are now
  differential invariants.  Find $t_q$, the number of independent
  functions over $M$ in $R^q$.
\item[6.] If $\dim G_q < \dim G_{q-1}$ or $t_q>t_{q-1}$, then increase
  $q$ by one, and go to step 2.
\item[7.] Otherwise, the algorithm terminates.  The differential
  invariants in $R^{q-1}$ furnish essential coordinates.   The
  isometry group has dimension $n-t_q+\dim G_q$. The orbits have
  dimension $n-t_q$.
\end{itemize}
\noindent
In principle, the essential coordinates obtained via the algorithm
allow the metric to be expressed in a canonical form that incorporates
the other differential invariants as essential constants and essential
functional parameters.

An analysis of the algorithm reveals that the conditions for an IC
order of $q=7$ are very stringent.  In \cite{coldinv93}, Collins and
d'Inverno list the following necessary conditions:
\begin{itemize}
\item[(C1)]  The components of the curvature tensor must be constants.
\item[(C2)] The invariance group at zeroth order $G_0$, must have dimension 2.
\item[(C3)] The dimension of the invariance group and the number of
  functionally independent components must not both change on
  differentiating.
\item[(C4)] We must produce at most one new functionally independent
  component on differentiating.
\item[(C5)] The dimension of the invariance group must go down by at
  most one dimension on differentiating.
\end{itemize}
Therefore, the search for a metric with $q=7$ must focus on a very
particular class of geometries.

To that end, let us say that a pseudo-Riemannian manifold is
\emph{curvature homogeneous} of order $k$, or $\CH_k$ for short, if
the components of the curvature tensor and its first $k$ covariant
derivatives are constant relative to an $\eta$-orthogonal frame.  Let
us also say that $M$ is \emph{properly} $\CH_k$ if it belongs to class
$\CH_k$, but is not a (locally) homogeneous space \cite{boeckx}.
Proper $\CH$ metrics are of central importance in invariant
classification, because they are the natural candidates for obtaining
sharp bounds on the IC order.

A homogeneous space can be characterized as a geometry without any
essential coordinates.  In other words, $t_q=0$; all differential
invariants are essential constants that define the structure of the
corresponding Lie algebra.  Thus, as a special case of the Karlhede
algorithm, we have the following result, first proved by
Singer \cite{singer} for Riemannian manifolds.
\begin{theorem}
   \label{thm:singcrit}
   If $(M,g)$ is a $\CH_{k+1}$ manifold and $\dim G_k= \dim G_{k+1}$, then
   $M$ is a (locally) homogeneous space.
\end{theorem}
Consequently, in a proper $\CH_k$ manifold, the first differential
invariant arises only at order $k+1$.  Thus, if only 1 new
differential invariant arises at each subsequent order, a proper
$n$-dimensional $\CH_k$ manifold could, in principle, have an IC order
$q=k+n+1$.  Below, we prove that this possibility can be realized with
$n=4$ and $k=2$.

We plan to report on a classification of proper $\CH_2$ and $\CH_1$
four-dimensional, Lorentzian geometries in a subsequent
publication \cite{milpel07}.  In principle, this will allow us to
investigate the sharpness of the IC bound for type D and O solutions,
as well.  In this letter, we focus on type N, and exhibit a family of
type N metrics that realize the $q=7$ bound.  The spacetimes in
questions belong to the class of null radiation, type N solutions on
an anti-de Sitter background.  This general class of exact solutions
was first investigated in \cite{DP81} and \cite{orr85}, but it was not
known that a $\CH_2$ geometry could arise as a special case.

Henceforth, $n=4$ and all tetrads $\{ \ve_a \} =
(\vm,\bvm, \vn,\vell)$ are complex, null-orthogonal with
\[ \bth_1 = \bth^2,\; \bth_3 = -\bth^4,\quad \bth^2 = \bar{\bth}^1,\;
\bth^3 = \bar{\bth}^3,\; \bth^4=\bar{\bth}^4 \] denoting the dual
coframe.  The metric is given by
\[g = 2\bth^1 \bth^2 - 2\bth^3 \bth^4.\]
The connection 1-form and the
the curvature 2-form 
are defined, respectively by
\begin{eqnarray}
  d\bth^a = \bom^a{}_b\wedge \bth^b,\quad \bom_{(ab)}=0\\
  \bOm^a{}_b = d\bom^a{}_b + \bom^a{}_c \wedge \bom^c{}_d.
\end{eqnarray}
The connection components are labeled by the 12 Newman-Penrose
scalars:
\begin{eqnarray}
  \label{eq:NPsc1}
  \qquad -\bom_{14}= \sigma\, \bth^1+\rho\, \bth^2+\tau\, \bth^3 +
  \kappa\, \bth^4;\\  
  \qquad \bom_{23} = \mu\, \bth^1+\lambda\,  \bth^2+\nu\, \bth^3
  +\pi\, \bth^4;\\ 
  \label{eq:NPsc3}
  -(\bom_{12}+\bom_{34})/2 = \beta\, \bth^1 +
  \alpha\,\bth^2+\gamma\,\bth^3+\epsilon\,\bth^4.
\end{eqnarray}
The curvature components are labelled by the Ricci
scalar $\Lambda=\bar{\Lambda}$, traceless Ricci components
$\Phi_{AB}=\bar{\Phi}_{BA},\; A,B=0,1,2$, and Weyl components $\Psi_C
,\; C=0,\ldots,4$:
\begin{eqnarray}
  \label{eq:NPcurv1}
  \small
  \fl \bOm_{14}= \Phi_{01}(\bth^{34}-\bth^{12})-
  \Phi_{02}\bth^{13}+\Phi_{00}\bth^{24} 
  +\Psi_0\bth^{14} -\left(\Psi_2 + 2\Lambda\right)\bth^{23}
  +\Psi_1 (\bth^{12}+\bth^{34})\\
  \label{eq:NPcurv2}
  \small
  \fl \bOm_{23} = \Phi_{21}(\bth^{12}-\bth^{34}) +\Phi_{22}
  \bth^{13}-\Phi_{20} \bth^{24}+\Psi_4\bth^{23} - 
  (\Psi_2 + 2\Lambda)\bth^{14}  -
  \Psi_3(\bth^{12}+ \bth^{34})\\
  \small
  \label{eq:NPcurv3}
  \fl (\bOm_{12} + \bOm_{34})/2=  - \Phi_{12}\bth^{13} +
  \Phi_{10}\bth^{24}+\Psi_1\bth^{14}-
  \Psi_3\bth^{23}+\\ \nonumber
  \small
  \qquad +\Phi_{11}
  (\bth^{34}-\bth^{12})   +(\Psi_2-\Lambda)(\bth^{12}+\bth^{34}),
\end{eqnarray}
where $\bth^{ab} = \bth^a\wedge\bth^b$.  Let
$\vA_\alpha=(A^a{}_{b\alpha})_{a,b=1}^n,\; \alpha=1,\ldots, n(n-1)/2$
be a basis of $\fo(\eta)$ with $C^\alpha{}_{\beta\gamma}$ the
corresponding structure constants. Let $\Gamma^\alpha{}_a,
R^\alpha{}_{ab}$, where
\begin{eqnarray}
  \bom^a{}_b = A^a{}_{b\alpha}\Gamma^\alpha{}_c \bth^c,\\
  \bOm^a{}_b = \frac{1}{2} A^a{}_{b\alpha} R^\alpha{}_{cd} \bth^{cd},
\end{eqnarray}
denote the connection and curvature components relative to the basis.

Let $\tLambda<0$ be a negative constant. Following \cite{orr85}, a
class of exact solutions for coupled electromagnetic radiation and
gravity propagating in a negatively curved background is given by
\begin{equation}
  \label{eq:orrmetric}
  \fl g_{ij} dx^i dx^j= 2p^{-2} d\zeta d\bar{\zeta} - 2 q^2 p^{-2}\,
  (( -(\tLambda A^2 + B\bar{B})r^2+ r\, q_s/q +
  2 H p/q) 
  ds + d r)d s,
\end{equation}
where $\zeta,\bar{\zeta},r,s$ are coordinates, where
\begin{eqnarray}
  p= 1+\tLambda \zeta \bar{\zeta},\\
  q = (1-\tLambda \zeta \bar{\zeta}) A +\bar{B}\zeta+B\bar{\zeta},\\
  A=A(s),\; B=B(s),\;   A=\bar{A},\\
  H =
  H(\zeta,\bar{\zeta}, s) ,\quad H_{\zeta\bar{\zeta}} + 2 \tLambda
  p^{-2} H = f\bar{f} p/q,
\end{eqnarray}
and where
\begin{equation}
  f \, d\zeta\wedge ds +\bar{f}\, d\bar{\zeta}\wedge ds, \quad f=f(\zeta,s),
\end{equation}
is the electromagnetic field.  

Let $\ttau_1\neq 0, \tmu_2\neq 8/5$ be real constants, and 
let $F(s)>0$ be a positive function of one variable.  Let us also set
\[ F_1(s) = F'(s)/F(s),\quad F_2(s) = (F_1'(s) -
F_1^2(s)/8)/\sqrt{F(s)},\] and demand that $F_2'(s)\neq 0$.  We
perform a change of coordinates and a specialization of the parameters
in \eqref{eq:orrmetric} as follows:
\begin{eqnarray}
  \fl \qquad\tLambda=-\ttau_1^2,\quad   A = 1,\quad B= -e^{3is} \ttau_1,\\ 
  \fl\qquad
  H =[36-72/p+(27+16\ttau_1^2 F(s)) q/p + 
  (10\tmu_2-16)p^3/q^3)]/(32\ttau_1^2) 
  \\ 
  \fl \qquad a = (\ttau_1/p) \,\Im(e^{-3is} \zeta),\\
  \fl \qquad b =
  \log(p)-\log(q)\\ 
  \fl \qquad t= r+a e^b(3/2+e^{2b}(1+4a^2/3))/\ttau_1^2.
\end{eqnarray}
The resulting metric can be expressed in terms of real coordinates
$a,b,s,t$  and a  null-orthogonal coframe as follows:
\begin{eqnarray}
  \label{eq:ch2om1}
  \fl\qquad \bth^1 = \left(db/2 + i(a\, db+ da)-e^b \left( a +
      i\left(a^2+\tmu_2/2-5/4\right) \right) ds
  \right)/\ttau_1,\\       
  \fl\qquad \bth^2 = \left(db/2 - i(a\, db+ da)-e^b \left( a -
      i\left(a^2+\tmu_2/2-5/4\right) \right) ds
  \right)/\ttau_1,\\       
  \fl\qquad  \bth^3 = e^b ds,\\
  \label{eq:ch2om4}
  \fl \qquad\bth^4 = e^{-3b}dt- (\tmu_2/\ttau_1^2)(da+a\, db) \\ \nonumber
  +\left(F(s)e^{-3b}- 6\,
    ae^{-2b}t 
    +(\tmu_2/\ttau_1^2) \left(a^2+\tmu_2/4-5/8\right)e^b \right)ds.
\end{eqnarray}
Using \eqref{eq:NPsc1}-\eqref{eq:NPcurv3} to calculate the connection
and curvature yields
\begin{eqnarray}
  \Lambda=-\ttau_1^2,\\
  \Phi_{22}=\Psi_4/3 = -4 + 5 \tmu_2/2,\\
  \tau=-\pi=2\beta=2\alpha/3=\ttau_1,\\
  \gamma= 3i/2,\\
  \mu=\lambda+2i = i \tmu_2,\\
  \nu=-3i\ttau_1 e^{-3b} t,
\end{eqnarray}
with all other NP scalars equal to zero.

As a basis of $\fo(\eta)$ let us take
\begin{equation}
   (\vA_\alpha) = (\ve_{23}, \ve_{13}, \ve_{34}, \ve_{12},
   \ve_{14}+\ve_{24}, \ve_{14} - \ve_{24}).  
\end{equation}
where $\ve_{ab}=\ve_a\wedge \ve_b$ is a basic bivector.  Note that,
because $\tmu_2\neq 8/5$, we have $\Phi_{22}, \Psi_4\neq 0$, and hence
the null rotations $\vA_5, \vA_6$ generate $G_0$.  Since the curvature
components are constant, we have
\begin{equation}
  \label{eq:DR1}
  \nabla_c R^\alpha{}_{ab}= \sum_{\beta=1}^4
  (\vA_\beta \cdot  R)^\alpha{}_{ab}\,\Gamma^\beta{}_c,
\end{equation}
where
\begin{equation}
  \label{eq:Raction}
  (\vA_\beta \cdot R)^\alpha{}_{ab} = C^\alpha{}_{\beta\gamma}
  R^\gamma{}_{ab} + 2 R^\alpha{}_{c[a} A^c{}_{b]\beta}
\end{equation}
denotes the standard action of $\fo(\eta)$ on the vector space of
curvature-type tensors.  Setting
\begin{equation}
  \label{eq:gamma1def}
  \Gamma^{(1)} = \left(\begin{array}{cccc}
    \sigma & \rho & \tau & \kappa \\
    \bar{\rho} & \bar{\sigma} &  \bar{\tau} & \bar{\kappa} \\
    -\beta-\bar{\alpha} & - \alpha-\bar{\beta} & -2\gamma_1 & - 2\epsilon_1\\
    -\beta+\bar{\alpha} & - \alpha+\bar{\beta} & -2i\gamma_2 & - 2i\epsilon_2
   \end{array}\right)
\end{equation}
where a subscript of 1 and 2 on a spin coefficient denotes the real
and imaginary part, respectively, lets us express the covariant
derivative, symbolically, as
\begin{equation}
  \label{eq:nablaR}
  \nabla R = \Gamma^{(1)} \cdot R.
\end{equation}
Since the components of $\Gamma^{(1)}$ are all constant, it follows
that the metric is $\CH_1$.  Next, since $\ttau_1\neq 0$, we have that
$\Gamma^{(1)}$, and thereby $\nabla R$, are invariant with respect
$\vA_6$. Thus, $G_1$ is generated by $\vA_6$. Setting
\begin{equation}
  \label{eq:gamma2def}
  \Gamma^{(2)}= \left( \begin{array}{c} \Gamma^{(1)} \\
    \frac{1}{2}(\mu+\bar{\lambda})   \quad
    \frac{1}{2}(\lambda+\bar{\mu})  \quad  
    \nu_1  \quad  \pi_1 
  \end{array}\right),
\end{equation}
and writing
\begin{equation}
  \label{eq:nabla2R}
  \nabla^{(2)}R = \Gamma^{(2)}\cdot \nabla R,
\end{equation}
we infer that the metric is $\CH_2$.  Note that $\Gamma^{(2)}$, and
thereby $\nabla^{(2)}R$ are not left invariant by $\vA_6$.  Hence,
$G_2$ is trivial, and
\begin{equation}
  \label{eq:nabla3R}
  \nabla^{(3)}R = \Gamma^{(3)}\cdot  \nabla^{(2)}R,
\end{equation}
where
\begin{equation}
  \label{eq:gamma3def}
  \Gamma^{(3)}= \left( \begin{array}{c} \Gamma^{(2)} \\
    \frac{1}{2}(\bar{\lambda}-\mu)  \quad \frac{1}{2}(\bar{\mu}-\lambda) \quad
    \nu_2 \quad \pi_2 
  \end{array}\right).
\end{equation}
Since $\nu_2$ is not a constant,  the metric is a
proper $\CH_2$ (not a $\CH_3$).  Indeed,
arising as a component of
$\nabla^{(3)} R$,
\begin{equation}
  \label{eq:I1def}
  I_1 := \nu_2 /(3\ttau_1) = e^{-3b} t
\end{equation}
is our first non-constant differential invariant.  
To calculate the frame derivatives, let us introduce the following
real coframe:
\begin{eqnarray}
  \label{eq:tth1def}
  \fl\quad \tth^1 = \ttau_1/2\, (\bth^1+\bth^2)\\
   \fl \qquad=db/2-ae^b ds,\\
   \fl\quad  \tth^2 = -i\ttau_1/2\, (\bth^1-\bth^2)+(2\tmu_2-5)/4\,
  \bth^3\\
  \fl \qquad =da+ a db- a^2 e^b\, ds,\\
    \fl\quad \tth^3 = \bth^3= e^b\, ds,\\
  \label{eq:tth4def}
   \fl\quad  \tth^4 = (\tmu_2/\ttau_1)(-i/2 \, (\bth^1-\bth^2)
  +(2\tmu_2-5)/(8\ttau_1)\, \bth^3) + \bth^4 \\
  \fl \qquad = e^{-3b}\left(dt+(F(s)-6ae^{b} t) ds\right)
\end{eqnarray}
Even though this coframe is not null-orthogonal, it's  form
simplifies the derivations that follow.  Since the two coframes are
related by a constant linear transformation, they yield the same
differential invariants.

Continuing, we have
\begin{equation}
  \label{eq:dI1}
   dI_1 = -6 I_1 \tth^1- e^{I_2} \tth^3 + \tth^4,
\end{equation}
where 
\begin{equation}
  \label{eq:I2def}
  I_2 := \log F(s) -  4b
\end{equation}
is our second differential invariant.  From the symbolic expression
\begin{equation}
  \label{eq:nabla4r}
  \nabla^{(4)} R = D \nabla^{(3)}R + \Gamma^{(3)} \cdot \nabla^{(3)} R
\end{equation}
we infer that $I_2$ can be recovered from the components of
$\nabla^{(4)} R$, and that  no other functionally independent invariants
occur in $\nabla^{(4)} R$. Next,
\begin{equation}
  \label{eq:dI2}
 dI_2 = -8\, \tth^1 +I_3\, \tth^3,
\end{equation}
where
\begin{equation}
  \label{eq:I3def}
  I_3 := e^{-b} F_1(s)-8a;
\end{equation}
and
\begin{equation}
  \label{eq:dI3}
   dI_3 = -2I_3\, \tth^1 -8\, \tth^2 +\left(I_3^2/8 + e^{I_2/2}
     I_4\right) \tth^3,  
\end{equation}
where
\begin{equation}
  \label{eq:I4def}
  I_4 := F_2(s).
\end{equation}
Hence, for analogous reasons we obtain $I_3$ from $\nabla^{(5)}R$ and
$I_4$ from $\nabla^{(6)}R$.  By assumption, $F_2(s)$ is non-constant,
and hence $I_1, I_2, I_3, I_4$ are functionally independent essential
coordinates. Since $n=4$, no functionally independent differential
invariants arise in $\nabla^{(7)}R$.  Indeed,
\begin{equation}
  \label{eq:dI4}
  dI_4 = F_2'(s) ds = e^{-b} F_2'(s) \tth^3 = e^{I_ 2/4} I_5 \, \tth^3,
\end{equation}
where
\begin{equation}
  \label{eq:I5de}
  I_5:= F_2'(s) (F(s))^{-1/4}.
\end{equation}
Since both $I_4$ and $I_5$ are functions of $s$, and since
$F_2'(s)\neq 0$,  locally
\begin{equation}
  \label{eq:phidef}
  I_5=\phi(I_4),
\end{equation}
where $\phi(x)$ is a function of 1 variable.
Therefore, the metric is fully classified by 2 essential constants
$\ttau_1, \tmu_2$ and by 1 essential functional parameter $\phi(x)$.
The above IC process can be summarized as follows:
\begin{equation}
  \label{eq:tsequence}
  (t_0,t_1,t_2,t_3,t_4,t_5,t_6,t_7) = (0,0,0,1,2,3,4,4).
\end{equation}
It is also possible to obtain the above count of differential
invariants directly, by means of a CAS package such as CLASSI \cite{aman}.
Therefore, the metrics described by
\eqref{eq:ch2om1}-\eqref{eq:ch2om4} have IC order $q=7$.  This proves
that the type N Karlhede bound is sharp.

\ack We gratefully acknowledge conversations with A. Coley, S.
Hervik, and G. Papadopoulos.  We also thank J. \AA man for useful
comments. The research of RM is supported in part by NSERC grant
RGPIN-228057-2004.


\section*{References}
\begin{thebibliography}{99}
\bibitem{aman} \AA man J E 2007 \textit{personal communication}
\bibitem{boeckx} Boeckx E, Kowalski O and Vanhecke L 1996,
  \textit{Riemannian manifolds of conullity two}, (River Edge, NJ:World
  Scientific)
\bibitem{brans65} Brans C H 1965 Invariant approach to the geometry of
  spaces in general relativity \textit{J. Math. Phys.} \textbf{6}
  95--102
\bibitem{cartan} Cartan E 1946 \textit{Le\c{c}ons sur la
    G\'{e}om\'{e}trie des Espaces de Riemann} (Paris:
  Gauthier-Villars)
\bibitem{coldinv93} Collins J M and d'Inverno R A 1993 The Karlhede
  classification of type-D non-vacuum spacetimes \textit{Class.
    Quantum Grav.}  \textbf{10} 343--51
\bibitem{DP81}  D\'{i}az A G and  Pleba\'{n}ski J F 1981 All nontwisting
  N's with cosmological constant 
  \textit{J. Math. Phys.} 2655--58
\bibitem{MacCAman86} MacCallum M A H and \AA man J E 1986
  Algebraically independent nth derivatives of the Riemann curvature
  spinor in a general spacetime \textit{Class. Quantum Grav.}
  \textbf{3} 1133--41
\bibitem{ramos98} Machado Ramos M P 1998 Invariant differential
  operators and the Karlhede classification of type N non-vacuum
  solutions \textit{Class. Quantum Grav.}  \textbf{15} 435--54
\bibitem{ramvick96} Machado Ramos M P and Vickers J A G 1996 Invariant differential operators and the Karlhede classification
  of type N vacuum solutions \textit{Class. Quantum Grav.} \textbf{13} 1589--99
\bibitem{karlhede80} Karlhede A 1980 A review of the geometrical
  equivalence of metrics in general relativity \textit{Gen. Rel.
    Grav.}  \textbf{12} 693--707
\bibitem{milpel07} Milson R and Pelavas N 2007 The curvature
  homogeneity bound for Lorentzian four-manifolds  \textit{Preprint}
\bibitem{orr85} Ozsv\'{a}th I, Robinson I and R\'{o}zga K 1985
  Plane-fronted gravitational and electromagnetic waves in spaces with
  cosmological constant \textit{J. Math. Phys} 1755--61
\bibitem{PSdI2000} Pollney D, Skea J E F and d'Inverno R A 2000
  Classifying geometries in general relativity: I. Standard forms for
  symmetric spinors \textit{Class. Quantum Grav.} \textbf{17} 643--63
\bibitem{PS2000} Paiva F M and Skea J E F 2000 On the invariant
  classification of conformally flat spacetimes \textit{IF/UERJ preprint}
  2000.002
\bibitem{singer} Singer I M 1960 Infinitesimally homogeneous spaces
  {\em Comm. Pure Appl. Math.} \textbf{13} 685--97.
\bibitem{skea2000} Skea J F 2000 A spacetime whose invariant
  classification requires the fifth covariant derivative of the
  Riemann tensor \textit{Class. Quantum Grav.}  \textbf{17} L69--L74
\end{thebibliography}
\end{document}